\documentclass[copyright,creativecommons]{eptcs}
\providecommand{\event}{HSB 2012}

\providecommand{\firstpage}{1}
\usepackage{graphics}
\usepackage{epsfig}
\usepackage{amsmath}
\usepackage{amssymb}
\usepackage{theorem}
\usepackage{ifthen}
\usepackage{subfig}

\title{On Expressing and Monitoring Oscillatory Dynamics}
\author{Petr~Dluho\v{s}, Lubo\v{s}~Brim, and David~\v{S}afr\'anek
\thanks{The work has been supported by the Grant Agency of Czech Republic grant GAP202/11/0312.}
\institute{Faculty of Informatics\\
Masaryk University\\
Botanick\'a 68a, Brno, Czech Republic}
\email{safranek@fi.muni.cz}
}
\def\titlerunning{On Expressing and Monitoring Oscillations}
\def\authorrunning{P.~Dluhos et al.}

\newtheorem{theorem}{Theorem}[section]
\newtheorem{definition}{Definition}[section]

\newtheorem{algorithm}{Algorithm}[section]
\newtheorem{remark}{Remark}[section]

\newenvironment{proof}[1][Proof]{\begin{trivlist}
 \item[\hskip \labelsep {\bfseries #1}]}{\end{trivlist}}

\newcommand{\df}{\overset{\text{df}}{=}}
\newcommand{\dfarrow}{\overset{\text{df}}{\Longleftrightarrow}}
\newcommand{\ax}[1]{{*\hspace*{-0.2em}\left[ #1 \right]}}
\newcommand{\until}[2]{\: \mathbf{U}_{
	\ifthenelse{\equal{#2}{}}{
    	#1}{
    	[#1, #2]}
    } \,
}
\newcommand{\future}[2]{\: \mathbf{F}_{ %mathcal
	\ifthenelse{\equal{#2}{}}{
    	#1}{
    	[#1, #2]}
    } \,
}
\newcommand{\globally}[2]{\: \mathbf{G}_{
	\ifthenelse{\equal{#2}{}}{
    	#1}{
    	[#1, #2]}
    } \,
}

\newcommand{\untilLTL}{\mathbf{U}}

\newcommand{\intervals}{\mathcal{J}}

\begin{document}
\maketitle

\begin{abstract}
To express temporal properties of dense-time real-valued signals, the Signal Temporal Logic (STL) has been defined by Maler et al. The work presented a monitoring algorithm deciding the satisfiability of STL formulae on finite discrete samples of continuous signals. The logic has been used to express and analyse biological systems, but it is not expressive enough to sufficiently distinguish oscillatory properties important in biology. In this paper we define the extended logic STL${}^*$ in which STL is augmented with a signal-value freezing operator allowing us to express (and distinguish) detailed properties of biological oscillations. The logic is supported by a monitoring algorithm prototyped in Matlab. The monitoring procedure of STL${}^*$ is evaluated on a biologically-relevant case study.
\end{abstract}

\section{Introduction}

In this paper we deal with automatic decision of the question if a given continuous signal satisfies a given temporal property. This question originally arose in the domain of analogous circuits verification~\cite{Maler_STL}. The procedure deciding this question has been called \emph{monitoring}~\cite{Maler_STL}. A monitor is constructed for a given property allowing off-line or even on-line decision over any continuous signal arising from a technical device (real case) or a numerical simulation procedure (model case)~\cite{MNP+08}. An important fact is that the monitoring procedure is always time bounded and that validity of the given logic property (discrete nature) is decided only approximately on a signal (continuous nature). These restrictions are necessary but not limiting to significantly help in avoiding errors in systems construction~\cite{AMT}.

In systems biology, continuous signals most typically represent the model case -- theoretical behaviour of mathematical models mimicking the dynamics of biological processes. Temporal properties are employed to express biological hypothesis~\cite{Pedro11, Calzone, BBS10, Paolo} and the monitoring procedure provides a promising analysis tool~\cite{BRJ+05, Rizk, Donze_Robustness}. Many dynamical phenomena that arise in biological systems have the form of oscillations. In physics the term oscillation represents an infinite behaviour periodically alternating certain quantities/qualities. The identifying aspect of oscillations is the fact that certain states in the phase space of the dynamical system are being repeatedly re-visited. A biological example of such phenomena are circadian rhythms. In~\cite{Nedbal} an interesting wet-lab experiment has been achieved on the cyanobacterium \emph{Cyanothece sp.} that showed a relation between metabolic cycles and circadian rhythms. A mathematical model of oscillations in gene regulation of cyanobacteria has been provided in~\cite{Miyoshi}. Other models targeting oscillatory behaviour can be found e.g. in~\cite{Kholodenko,Elowitz_Repressilator}. 

In contrast to physics, the notion of oscillation in biology is usually understood informally and in a wider scope. Many wet-lab experiments show oscillatory behaviour with decreasing amplitude (so-called \emph{damped oscillation}). Dual notion, oscillation with increasing amplitude, is also relevant: the question targeting how many oscillations and how strongly they increase until a permanent oscillation is achieved comes to a significant interest when tuning biological processes via mathematical models~\cite{Kholodenko,Elowitz_Repressilator}. When considering the population behaviour as a real-valued signal in dense-time domain, we need to quantitatively express and compactly encode the mentioned types of oscillatory phenomena.

In~\cite{Calzone}, a linear-time logic with constraints over real values has been defined. It can be practically used to express many dynamical phenomena including oscillations of species concentration. The logic is interpreted over finite time-series of real-valued data obtained as discretely sampled solutions of differential equations (ODEs) (or a series of physical measurements). Formally, a technical problem may arise with the dense-time domain of considered signals. In general, in a particular interval on the time axis of the signal there may exist infinitely many points where a considered property (a predicate over real-values) changes its truth value. A rigorous semantics which treats this problem is given in~\cite{Maler_STL} in terms of continuous (real-valued) signals which are required to be piece-wise affine. In such a setting, it is assumed that there exists a finite interval covering of the signal time-domain where the end points can be sampled and on which the signal behaves ``reasonably'' in each individual time interval. The logic with such a semantics is called \emph{Signal Temporal Logic (STL)}~\cite{Maler_STL} and is based on the (bounded) Metric Interval Temporal Logic (MITL)~\cite{Alur_MITL}.

\begin{figure}[ht]

	\centering

		\subfloat[][$x(t) = sin(\frac{2\pi t}{10})$]{\includegraphics[scale=0.2]{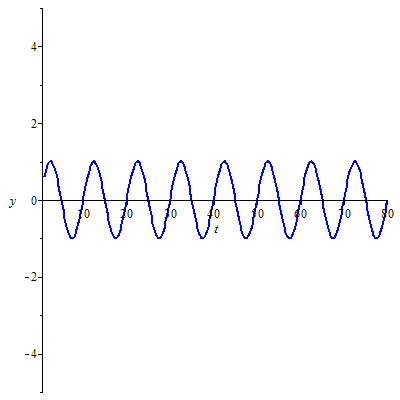}
			\label{figOscillationsClasses-a}}\hspace{1cm}
		\subfloat[][$x(t) = \frac{t}{20}sin(\frac{2\pi t}{10})$]{\includegraphics[scale=0.2]{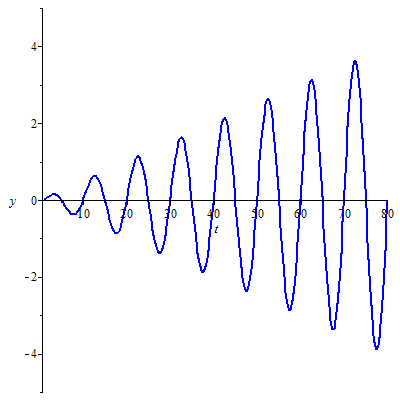}\hspace{1cm}
			\label{figOscillationsClasses-b}}
		\subfloat[][$x(t) = \frac{20}{t}sin(\frac{2\pi t}{10})$]{\includegraphics[scale=0.2]{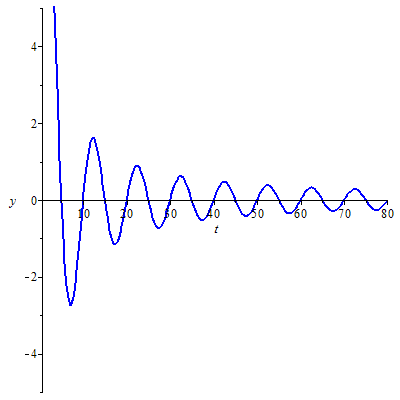}\hspace{1cm}
			\label{figOscillationsClasses-c}}\\

	\caption{Various types of oscillations.}

	\label{figOscillationsClasses}

\end{figure}

Neither STL, nor other temporal logic, as far as we know, provide any possibility to express (and distinguish) the classes of oscillations such as damped oscillations or oscillations with increasing amplitude. The reason is impossibility of globally referencing (and relatively comparing) concrete signal values occurring in time points in which some local property is satisfied. Of course, in STL we can express directly any single concrete signal, but not a universal property without references to concrete values. E.g., the damped oscillation in Figure~\ref{figOscillationsClasses-c} can be expressed in STL as a sequence exactly reaching the 15 local extremes in the given order. In general, there can appear any number of local extremes in the observed time interval. Such a general property cannot be expressed in STL. 

In this paper, we propose an extension of STL, denoted STL$^*$, based on enriching the logic with a \emph{signal-value freeze operator} $*[\varphi]$ which allows referencing to a signal value in a time point when $\varphi$ is determined true. By means of atomic propositions enriched by variables of the form $x^*$ referring to ``frozen'' values, we can express damped oscillations by the following formula:

%One could think of more different formulae describing the class of oscillatory signals and they could vary in some marginal cases. This depends on what exactly do we accept as the desired behaviour.
%A typical phenomenon in biological systems is a \textit{damped oscillation}. It is an oscillation with gradually decreasing amplitude (Figure \ref{figOscillationsClasses-c}). This property can be expressed by the formula:

%\begin{equation}

%	\label{propertyDampedOscillation}

%$$\globally{0}{60} ((\future{0}{10} \ax{\globally{0}{10} x^\ast \geq x}) \wedge (\future{0}{10} \ax{\globally{0}{10} x^\ast \leq x})).
%$$

$$\globally{0}{60} ((\future{0}{10} \ax{\globally{1}{10} x^\ast \geq x + c}) \wedge (\future{0}{10} \ax{\globally{1}{10} x^\ast \leq x - c})).$$
%\end{equation}

It says that \textit{"there is always a time instant in near future which is a local maximum for some future interval and there is also another time instant in near future which is a local minimum for some future interval"}. In these intervals, another local maxima and minima for more distant intervals occur, and so on. This implies that the signal values can be surrounded by a decreasing function from top and an increasing function from bottom. The constant $c$ sets the extent of damping.

%Among the signals in Figure \ref{figOscillationsClasses}, only the signals a) and c) satisfy the formula \eqref{propertyDampedOscillation}. The signal a) is a marginal case of damped oscillation with zero damping. We cannot get rid of this case by simply substituting strict inequalities for the nonstrict in \eqref{propertyDampedOscillation} because the condition $\globally{0}{10} x^\ast > x$ would never hold for $x^\ast = x$. However, we could shift the left point of the interval and add a constant $c > 0$ to the formula \eqref{propertyDampedOscillation} to ensure the damping would proceed with some speed:

%\begin{equation}

%	\label{propertyDampedOscillation2}

%$\globally{0}{60} ((\future{0}{10} \ax{\globally{1}{10} x^\ast \geq x + c}) \wedge (\future{0}{10} \ax{\globally{1}{10} x^\ast \leq x - c})).$

%\end{equation}

%Another class of properties which cannot be expressed in STL are the relations between variables in different times. An example is the property \textit{"$x$ copies the values of $y$ with a delay of 4 seconds"} or equally \textit{"$x$ equals to the value $y$ had 4 seconds ago"}. In STL*, it corresponds to the formula:

%\begin{equation}

%	\label{property_zpozdeni}

%$\globally{0}{80 - (4 + \delta)} \ax{\globally{4 - \delta}{4 + \delta} y^\ast = x},
%$
%\end{equation}

%for a small $\delta > 0$ (recall temporal operators can be bounded only by nonsingular intervals. It is due to the same restriction in MITL, which is a necessary condition for MITL to be decidable~\cite{Alur_MITL}).

% Future ... Monitoring of  wet-lab measurements 

The concept of freeze quantification has been introduced in~\cite{Alur_TPTL} to increase real-time temporal logic expressiveness by allowing every temporal operator to bind a time variable to the time it refers to. In the context of systems biology this concept has been used in Biological Oscillator Synchronisation Logic (BOSL)~\cite{Bartocci} to reason on oscillators synchronisation. In our work we shift the notion of temporal quantification to bind the real-valued signal variables to the temporal reference of time and we do it at the level of STL logic. Freezing of variable values has been used before in higher-level formalisms employing local variables~\cite{EF08,SystemVerilog}. However, these formalisms do not support dense-time binding.

It is necessary to note that when values of derivatives of signal values are known in all sampled points, some oscillatory properties can be expressed by using plain STL with predicates over derivatives. However, in such a case all needed derivatives (of all required orders) must be included within the signal. Getting these data can be computationally hard or even impossible. STL$^*$ allows to express (and monitor) oscillatory properties without the need of any additional information supporting the signals.

The contribution of this paper is an extension of STL logic with signal value freeze quantification (Section~\ref{sec:stlstar}) motivated by the need to express oscillations appearing in biology. The paper is primarily focused on a practical aspect, we give an algorithm for monitoring STL$^*$ formulae over (bounded) continuous signals (Section~\ref{sec:algorithm}). The result is supported by a prototype implementation evaluated on a biological case study of \emph{E. coli} repressilator (Section~\ref{sec:casestudy}). Please note that a detailed discussion of STL$^*$ expressiveness is provided in a master's thesis~\cite{Dluhos} that makes a preliminary version of this paper.

\section{Background}
%\subsection{Signals}

First, we briefly recall the real-valued signals over continuous time from~\cite{Maler_STL}. For the purposes of this paper, we focus only on signals over $\mathbb{R}^n$.% [TODO: define on finite closed intervals? - STL time domains are right-open intervals]

\begin{definition}
	\label{defSignal} Let $T = [0, r], r \in \mathbb{Q}_{\geq0}, T \subseteq \mathbb{R}_{\geq0}$, be a finite interval (\emph{time domain} of the signal). A \emph{finite length signal $s$} is a function $s : T \rightarrow \mathbb{R}^n$ where $|s| = r$ denotes the \emph{length of the signal $s$}. 
\end{definition}

\begin{definition}
	Given a signal $s : T \rightarrow \mathbb{R}^n$, $n$ denotes the \emph{order of the signal $s$}. We use $s_i$ to denote the $i$-th component of the signal, $s_i : T \rightarrow \mathbb{R}$ (i.e., the canonical projection to the $i$-th element).
	%, defined as $s_i(t) \df \pi_i(s(t))$ where $\pi_i$ is the projection from $n$-tuples to the $i$-th element.
\end{definition}

A signal of order $n$ represents one finite run of a continuous-time system with $n$ variables. A variable can represent any quantity of the system, e.g., concentration of species, its derivative or any other attribute of the signal which can be measured and quantified.

To translate the properties of a signal into terms of logic, we use a function transforming real values of the signal components in every time instant to the Boolean set $\lbrace 0, 1 \rbrace$, representing satisfaction or dissatisfaction of a property over the signal. These truth values are used as propositional variables in formulae of the logic. Unlike in~\cite{Maler_STL}, we restrict the allowed operations on signal variables to linear combinations and simple comparisons. Such a restriction is needed for our monitoring algorithm (Section \ref{SectionAlgorithm}) and does not limit expression of the properties of our interest.

%prevent expressing studied phenomena.  
  
%We need a formalism to express properties of a signal in terms of logic. To do so, we define a function transforming real values of the signal variables in every time instant to the Boolean set $\lbrace 0, 1 \rbrace$, representing satisfaction or dissatisfaction of one property over one signal. This truth values can be then used as propositional variables in some logic.

%We will restrict the allowed operations on signal variables to linear combinations and simple comparisons. Such restriction enables us to algorithmically determine satisfaction of a property of a signal in every time instant. By using better algorithms or less precise solution we could loosen the demands on the form of the properties.
%TODO: shoud be discussed somewhere?

\begin{definition}
	\label{defLinearPredicate} \emph{Linear predicate $\nu$ of order $m$} is a function $\nu : \mathbb{R}^m \rightarrow \mathbb{B}$ of the form $\sum_{i=1}^{m} a_i x_i \sim b$ where $a_i, b \in \mathbb{R}$ are real coefficients, $x_i$ are input variables, $\sim \in \lbrace <, \leq, >, \geq, = \rbrace$ and $\mathbb{B}$ is a Boolean set $\mathbb{B} = \lbrace 0, 1 \rbrace$. The predicate $\nu(x_1, x_2, \ldots x_m)$ returns $1$ iff the expression $\sum_{i=1}^{m} a_i x_i \sim b$ is true and $0$ otherwise.
\end{definition}

However, in the logic defined in this paper, we need not only to work with the values of variables acquired in a single time instant, but also with a second set of values from another time instant. For this reason, we will define an additional function extracting truth values from the signal.
 
\begin{definition}
	\label{defPredicate} \emph{Atomic predicate $\mu$} is a function $\mu : \mathcal{S}_n \times \mathbb{R}_0^+ \times \mathbb{R}_0^+ \rightarrow \mathbb{B}$ where $\mathcal{S}_n$ is a class of signals of order $n$. For each signal $s \in \mathcal{S}_n$ and $t, t^\ast \in [0, |s|]$, $\mu$ is defined by the formula: $$\mu(s, t, t^\ast) \df \nu(s_1(t), \ldots, s_n(t), s_1(t^\ast), \ldots, s_n(t^\ast)),$$ where $\nu$ is a linear predicate of order $2n$ and $s_i$ are components of the signal $s$. Values of the atomic predicate $\mu$ for $t > |s|$ or $t^\ast > |s|$ are considered undefined.% because the signal $s$ is undefined for these time instants.
\end{definition}

An atomic predicate over a signal $s$ describes a Boolean property of the signal $s$, i.e., if a constraint specified by the linear predicate $\nu$ is satisfied with respect to two time instants $t$ and $t^\ast$. An example of such a property is \textit{"under condition $t^\ast < t$, the difference between values of variables $s_1$ and $s_2$ was higher in the time $t^\ast$ than in the time $t$"}:
\begin{equation*}
	\mu(s, t, t^\ast) = \nu(s_1(t), s_2(t), s_1(t^\ast), s_2(t^\ast)) = s_1(t) - s_2(t) < s_1(t^\ast) - s_2(t^\ast),
\end{equation*}
which can be rearranged to the standard form of a linear predicate according to Definition \eqref{defLinearPredicate}:
\begin{equation*}
	\nu(s_1(t), s_2(t), s_1(t^\ast), s_2(t^\ast)) \df s_1(t) - s_2(t) - s_1(t^\ast) + s_2(t^\ast) < 0.
\end{equation*} 

Instead of $\mu(s, t, t^\ast) = s_1(t) - s_2(t) < s_1(t^\ast) - s_2(t^\ast)$ we usually write $\mu(s, t, t^\ast) = x - y < x^\ast - y^\ast$, denoting the variables of a signal $s$ in time $t$ by letters $x, y, z, \ldots$ and the same variables in time $t^\ast$ as $x^\ast, y^\ast, z^\ast, \ldots$.

As in~\cite{Maler_STL}, we will avoid the undesired cases of atomic predicates where the output values of the predicate are varying infinitely often. %From now on, we will deal only with signals which values, with respect to all used atomic predicates, changes at most finitely often. It is a reasonable presumption because the signals of our interest are usually outputs of some systems measured in finite-length time steps. Such a pathological behaviour would vanish during this process of sampling and cannot reappear due to linear form of atomic predicates.
We assume that we
deal with signals that are well-behaving with respect to every $\mu$, that is,
$\mu(s)$ has a bounded variability and every change in $\mu(s)$ is detected in the
sense that every point t such that $\mu(s[t]) \neq lim_{t'->t} \mu(s[t'])$ is
included in the sampling~\cite{Maler_STL}.

\section{A Logic for Oscillatory Dynamics}
\label{sec:stlstar}

In this section we describe the syntax and semantics of STL*. It is based on the Signal Temporal Logic (STL)~\cite{Maler_STL} which we extend with the signal-value freeze operator $*[\varphi]$. We consider atomic predicates restricted to linear as defined in the previous section.

\subsection{Syntax}
The formulae of the STL* are inductively defined by the following grammar:
 
\begin{equation*}
		\varphi ::= \mu_i \mid \neg \varphi \mid \varphi_1 \vee \varphi_2 \mid \varphi_1 \until{a}{b} \varphi_2 \mid \ast \varphi, \\
\end{equation*}

where $\mu_i$ are atomic predicates from Definition \ref{defPredicate}, $\neg$ and $\vee$ are standard propositional logic operators, $\until{a}{b}$ is a bounded until operator constrained by the closed nonsingular time interval $[a, b]$ with rational end-points. The interpretation of this operator is similar to the one used in the Metric Interval Temporal Logic (MITL)~\cite{Alur_MITL}. Finally, $\ast$ is the newly added unary \emph{signal-value freeze operator}. In the standard way, we can derive additional temporal operators such as eventually ($\future{a}{b} \varphi \equiv \mathsf{true} \until{a}{b} \varphi$) and globally ($\globally{a}{b} \varphi \equiv \neg \future{a}{b} \neg \varphi$). %These operators are then also bounded by an interval $[a, b]$ and their interpretation follows from the interpretation of $\until{a}{b}$.\\ 

\subsection{Semantics}
Formulae of STL* are interpreted over triplets $(s, t, t^\ast)$ where $s$ is a real-valued signal and $t, t^\ast \in [0, |s|]$ are two time instants. We write
%\begin{equation*}
	$(s, t, t^\ast) \models \varphi$,
%\end{equation*}
iff a formula $\varphi$ is satisfied on the triplet $(s, t, t^\ast)$. The satisfaction of an STL* formula is defined inductively to its structure:

\begin{equation}
	\label{STL*_sematics}
    \begin{array}{lcl}
    	(s, t, t^\ast) \models \mu & \dfarrow & \mu(s, t, t^\ast) = 1; \\
    	(s, t, t^\ast) \models \neg \varphi & \dfarrow & (s, t, t^\ast) \not \models \varphi; \\
    	(s, t, t^\ast) \models \varphi_1 \vee \varphi_2 & \dfarrow & (s, t, t^\ast) \models \varphi_1 \: \textnormal{or } \: (s, t, t^\ast) \models \varphi_2;  \\
    	(s, t, t^\ast) \models \varphi_1 \until{a}{b} \varphi_2 & \dfarrow & \exists t' \in [a + t, b + t]: (s, t', t^\ast) \models \varphi_2 \, \wedge  \\
    	& & \forall t'' \in [t, t']: (s, t'', t^\ast) \models \varphi_1;\\
    	(s, t, t^\ast) \models \ast \varphi & \dfarrow & (s, t, t) \models \varphi. \\
	\end{array}
\end{equation}
Because models for STL* are signals, we also speak about satisfaction of a formula over a signal.

\begin{definition}
\label{defSatisfactionSignal}
The satisfaction of a formula $\varphi$ over a signal $s$ is defined as:
%\begin{equation*}	
	$s \models \varphi \dfarrow (s, 0, 0) \models \varphi$.
%\end{equation*}
\end{definition}

The meaning of the operator $\ast \varphi$ is to freeze the values of the signal over which the formula is interpreted in the current time point so we can use these values inside the subformula $\varphi$. Formula $\ast \varphi$ is true in time $t = t_0$ iff the formula $\varphi$ is true with the time $t^\ast$ frozen in this time, i.e., $t^\ast = t_0$. In combination with other temporal operators, interesting properties can be expressed such as \textit{"the value of the variable $x$ is nondecreasing on the interval I"},
%\begin{equation*}
	$\varphi_1 \equiv \globally{I}{} \ax{ \globally{0}{\epsilon} (x^\ast \leq x) }$,
%\end{equation*}
where $\epsilon > 0$ is an arbitrary constant. Or \textit{"at some point in the future (during interval $I$) the value x increases by the value of 5 within two time units"},
%\begin{equation*}
	$\varphi_2 \equiv \future{I}{} \ax{ \future{0}{2} (x^\ast + 5 = x) }$.
%\end{equation*}

To determine the satisfaction of a formula over a signal, the signal has to be of a sufficient length. The necessary length can be computed for each formula inductively to its structure as in~\cite{Maler_STL}. The freeze operator does not require any additional length:
%Determination of satisfaction of a formula $\varphi$ over a signal $s$ for $t > |s|$ or $t^\ast > |s|$ is impossible due to lack of information about values of atomic predicates at these time instants. This follows from the finite nature of considered signals and it is also the reason why the temporal operator $\until{I}{}$ is bounded only by constrained interval. To avoid problems with undefined values, we can compute the sufficient length $l(\varphi)$ of a signal inductively according to the structure of the interpreted formula $\varphi$:

\begin{equation}
	\label{sufficientSignalLength}
    \begin{array}{lcl}
    	l(\mu) & \df & 0;\\
    	l(\neg \varphi) & \df & l(\varphi);\\
    	l(\varphi_1 \vee \varphi_2) & \df & \textnormal{max}(l(\varphi_1), l(\varphi_2));\\
    	l(\varphi_1 \until{a}{b} \varphi_2) & \df &  \textnormal{max}(l(\varphi_1), l(\varphi_2)) + b;\\
    	l(\ast \varphi) & \df & l(\varphi).
	\end{array}
\end{equation}

From Definition \ref{defSatisfactionSignal} we can see that if a formula $\varphi$ contains an atomic predicate $\mu_i$ and there is no operator $\ast$ wrapping this predicate then all variables concerning the frozen time instant $t^\ast$ are handled as if $t^\ast = 0$. 

From the definition of the semantics of STL* \eqref{STL*_sematics} follows that in a formula with several nested operators $\ast$, the meaning of a frozen variable $x^\ast$ is local. I.e., it relates to the nearest operator $\ast$, because the content of the variable $t^\ast$ is overwritten by the most nested freeze operator $\ast$. It implies that only a single set of frozen signal values is accessible at every place in the formula. For example, in a formula:
\begin{equation}
\label{property2}
\ax{ x^\ast \leq y \until{I}{} (x^\ast = y \wedge \ax{ \globally{0}{5} x^\ast \geq y } ) },
\end{equation}
the first and the second occurrence of the variable $x^\ast$ relate to the first usage of $\ast$ and address the time $t = 0$, in which the formula is interpreted. The third occurrence of $x^\ast$, in contrast to the previous two, relates to the second operator $\ast$ and addresses the time instant in which the subformula on the right side of $\until{I}{}$ becomes true. Semantics restricted in this way does not excessively limit the capability of expressing various biological behaviour while remaining computationally feasible.% Allowing more coexisting freezing points would expand the complexity of the monitoring algorithm presented in Section \ref{SectionAlgorithm}.

%The property described by the formula \eqref{property2} can be roughly expressed as \textit{"when $y$ becomes larger than the actual value of $x$ on the interval $I$ for the first time (say at time $t = t_0$), it will not fall under the value of $x(t_0)$ for the next 5 time units"}.\\

\section{Algorithm}
\label{SectionAlgorithm}
\label{sec:algorithm}

In this section we deal with monitoring of an STL* formula, which means to determine the satisfaction of a formula over a finite length signal. 
%MOVE TO INTRO
%Unlike model checking (i.e., determining if a formula is satisfied by all runs of a system~\cite{Clarke_ModelChecking, ModelChecking_book}), monitoring is much simpler task~\cite{Maler_STL}. It does not guarantee the satisfaction over the whole space of solutions of a system, but it can be performed repeatedly to sample the searched space to increase the knowledge about the system. It can be also used for parameter estimation [TODO: reference]
The monitoring procedure introduced in this paper is inspired by constrained LTL model-checking~\cite{Calzone} and monitoring of STL formulae~\cite{Maler_STL}, but needed to be extended into 2D space due to freeze operator $\ast$. It is because the task is solved simultaneously for two time instants, actual time and frozen time.

The idea of the monitoring procedure is to construct a parse tree of the formula and check the satisfaction in a bottom-up manner. First step in checking a formula $\varphi$ over a signal $s$ is to construct a set of time points in which the formula $\varphi$ is satisfied.

\begin{definition}
	Let $\varphi$ be a STL* formula and $s$ be a real-valued signal. A~set $S_{\varphi,s} = \lbrace (t, t^\ast) \in [0, |s|] \times [0, |s|] \mid (s, t, t^\ast) \models \varphi \rbrace$ is called the \emph{satisfaction set} of formula $\varphi$ over signal $s$. We use short-form notation $S_\varphi$ whenever the signal $s$ is obvious from the context.
\end{definition}

Now we inductively construct satisfaction sets for nodes on higher levels of the parse tree. The last step of the procedure is to decide if $s \models \varphi$ from satisfaction set for the formula $\varphi$. According to Definition \ref{defSatisfactionSignal}, it is equivalent to checking whether the satisfaction set contains the point $(0, 0)$.

Constructing the satisfaction set for a general signal can be a very difficult task. For this reason we will consider the monitoring algorithm only for piecewise linear signals. This is a reasonable requirement because in most cases we deal with time series produced by numerical simulations of modelled systems or by some measurements. These series can be interpreted as piecewise linear signals considering the values changing linearly between two adjacent points. If required, other points can be generated in between the existing points to make the signal more precise.

\begin{definition}
	A real-valued signal $s$ of order $n$ is called \emph{piecewise linear signal} iff all the projections $s_i(t) : [0, |s|] \rightarrow \mathbb{R}, i = 1, \ldots, n$ are piecewise linear functions defined on a finite set of intervals $\intervals = \lbrace I_1, I_2, \ldots, I_m \rbrace$ where $I_j \subseteq [0, |s|],j = 1,\ldots, m$ and $\bigcap\intervals= \emptyset, \bigcup\intervals= [0, |s|]$.
\end{definition}

\begin{theorem}[Inductive construction of the satisfaction set]
	\label{monitoring_STL*}
	Let $s$ be a piecewise linear signal and $\varphi$ a formula in STL*. The satisfaction set $S_\varphi$ of the formula $\varphi$ over the signal $s$ can be constructed inductively with respect to the structure of the formula $\varphi$:       
	\begin{equation*}
    	\begin{array}{lcl}
        	S_\mu & = & \lbrace (t, t^\ast) \in \mathbb{R}_0^+ \times \mathbb{R}_0^+ \mid \mu(s, t, t^\ast) = 1 \rbrace;\\
    		S_{\neg \varphi} & = & \mathbb{R}_0^+ \times \mathbb{R}_0^+ \setminus S_\varphi;\\ 
    		S_{\varphi_1 \vee \varphi_2} & = & S_{\varphi_1} \cup S_{\varphi_2};\\
			S_{\varphi_1 \until{a}{b} \varphi_2} & = & \lbrace (t, t^\ast) \in S_{\varphi_1} \mid \exists t' \in [t + a, t + b]: (t', t^\ast) \in S_{\varphi_2} \, \wedge\\
			& & \forall t'' \in [t, t'] : (t'', t^\ast) \in S_{\varphi_1}\rbrace;\\
			S_{\ast \varphi} & = & \lbrace (t, t^\ast) \in \mathbb{R}_0^+ \times \mathbb{R}_0^+ \mid (t, t) \in S_{\varphi}, t^\ast \in \mathbb{R}_0^+ \rbrace.\\ 
		\end{array}
	\end{equation*}

	\begin{proof}
		Each of the equations follows directly from the definition of semantics of STL* \eqref{STL*_sematics}.
	\end{proof}
\end{theorem}

The assumption of piecewise linear signals in combination with linearity of atomic predicates (Definitions \ref{defLinearPredicate}, \ref{defPredicate}) enables us to construct satisfaction sets for atomic predicates in polynomial time and to easily compute sets belonging to higher formulae. The formalism we use for the representation of satisfaction sets are convex polytopes.

\begin{definition}
	\label{defConvexPolytope}
	\emph{Convex polytope} $P \subseteq \mathbb{R}^n$ is a set $\lbrace x = (x_1, x_2, \ldots, x_n) \in \mathbb{R}^{n \times 1} \mid Ax \leq b \rbrace$ where $A \in \mathbb{R}^{m \times n}$ is a matrix and $b \in \mathbb{R}^{m \times 1}$ is a  column. A convex polytope in $\mathbb{R}^1$ is called a \emph{line segment}, in $\mathbb{R}^2$ a \emph{convex polygon}.
\end{definition}

Each inequality $\sum_{j=1}^n a_{ij} x_j \leq b_i$ in Definition \ref{defConvexPolytope} identifies a subspace of $\mathbb{R}^n$. The convex polytope is obtained as intersection of these subspaces.

\begin{figure}[h]
\vspace*{-3mm}
	\begin{center}
    \includegraphics[scale=0.4]{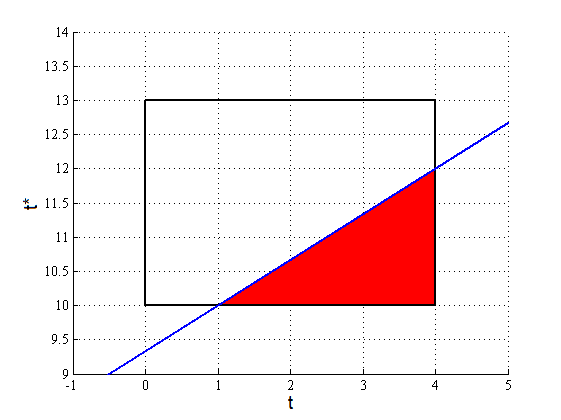}
     \caption{An illustration of the set $S^+_{i,j}$ for a formula $x^* < \frac{2}{3}x + \frac{28}{3}$ on the subspace $I_i \times I_j = [0, 4] \times [10, 13]$.}
    \label{figCut}
    \end{center}
\vspace*{-3mm}
\end{figure}

\begin{theorem}
	\label{lemma}
	Assume $s$ a piecewise linear signal and $\intervals = \lbrace I_1, \ldots, I_m \rbrace$ the set of intervals on which $s$ is defined. Let $\mu$ be an atomic predicate over the signal $s$. The predicate $\mu(s, t, t^\ast) = \sum_{i=1}^{n} a_i s_i(t) + \sum_{i=1}^{n} b_i s_i(t^\ast) \sim b, \sim \in \lbrace <, \leq, >, \geq, = \rbrace$ is a linear function over the signal $s$ in variables $t$ and $t^\ast$ according to Definition \ref{defPredicate}. Denote $S^+_{i, j} \df \lbrace (t, t^\ast) \in I_i \times I_j \mid \mu(s, t, t^\ast) = 1 \rbrace$ where $i,j \in \lbrace 1, \ldots m \rbrace, I_i, I_j \in\intervals$ the set of time instants at which the atomic predicate $\mu$ is satisfied over signal $s$ on rectangular area $I_i \times I_j$ and $S^-_{i,j} \df (I_i \times I_j) \setminus S^+_{i,j}$ its complement. For $t \in I_i, t^\ast \in I_j$ there can arrive two situations:

\begin{enumerate}
	\item if $\sim\,\equiv\,=$, then the set $S^+_{i,j}$ makes a line segment (possibly degenerated to a single point or an empty set); 
	\item if $\sim\,\not\equiv\,=$, then the space $I_i \times I_j$ is divided by the line $\sum_{i=1}^{n} a_i s_i(t) + \sum_{i=1}^{n} b_i s_i(t^\ast) = b$ into two subspaces $S^+_{i,j}$ and $S^-_{i,j}$ (Figure \ref{figCut}). (Again, there might be a degenerated case when the line $\sum_{i=1}^{n} a_i s_i(t) + \sum_{i=1}^{n} b_i s_i(t^\ast) = b$ does not cross the rectangle $\lbrace (t, t^\ast) \in I_i \times I_j \rbrace$. In this case $S^+_{i,j} = \emptyset$ and $S^-_{i,j} = I_i \times I_j$ or vice versa).
\end{enumerate}

\begin{proof}
		Described properties result from the linear form of atomic predicates \ref{defPredicate} and from the fact that signal $s$ behaves linearly on each interval $I_i \in\intervals, i \in \lbrace 1, \ldots m \rbrace$.
	\end{proof}

\end{theorem}

In both cases, the set $S^+_{i,j}$ makes a convex set easily describable by set operations over convex polytopes. In the first case, a line segment itself is a convex polytope. %(two inequalities $\sum_{i=1}^n a_i x_i \leq b$ and $-\sum_{i=1}^n a_i x_i \leq -b$ degrade one dimension into a line and a line segment can be acquired by intersection with the rectangle $I_1 \times I_2$), 
In the second case, either the set $S^+_{i,j}$ is a convex polytope, or (when $\sim \in \lbrace <, > \rbrace$) it is a convex polytope after subtraction of one bordering line segment, which is a polytope.

Using this algorithm we can express the satisfaction set $S_{\mu}$ as $S_{\mu} = \bigcup_{i,j \in\{1,...,m\}} S^+_{i,j}$, where $S^+_{i,j}$ are the sets described above constructed by set operations on polytopes. The set $S_{\mu}$ can be further simplified by joining adjacent polytopes. Having satisfaction sets for atomic predicates, we could inductively construct satisfaction sets for composite formulae from Theorem \ref{monitoring_STL*} again as operations on polytopes, which could be achieved in polynomial time~\cite{GeometricAlgorithms}. 

However, working with precise satisfaction sets as defined in Theorem \ref{monitoring_STL*} can be quite a difficult task. We have to work with polytopes of different dimensions (polygons, lines and points) and operations over sets of these objects can be unnecessarily demanding. In real-world applications, we could end up performing an expensive monitoring over noisy signals measured on imperfect devices or computed on computers with finite numerical precision. Hence think about less precise, yet faster monitoring algorithm.

%%%%%%%%%%%%%% approximative algorithm %%%%%%%%%%%%%%%%%%%

We can imagine a \emph{noisy signal} as a signal measured or computed with finite precision. The actual values of the variables can differ from the values presented by the signal up to some error $\epsilon > 0$. This inaccuracy in the domain of values implies also inaccuracy in the time domain. When investigating the time instant in which some event occurred (e.g., $x > 0$) we cannot be sure when exactly it happened. That is because the critical value $x = 0$ is burdened by an error and it is possible that $x = 0$, but also that $x > 0$ or $x < 0$ in that particular time instant.

%Suppose we have a noisy piecewise linear signal $s$ measured or computed with some error $\epsilon > 0$, let $\intervals$ be the set of intervals on which $s$ behaves linearly. Let $\mu(s, t, t^\ast)$ be an atomic predicate over the signal $s$. For $t \in I_i, t^\ast \in I_j; I_i, I_j \in\intervals$ there could arrive two situations as discussed in Theorem \ref{lemma}:

%\begin{enumerate}
%	\item if $\sim \, \equiv \, =$, then the set $S^+_{i, j}$ makes a line segment; 
%	\item if $\sim \, \not\equiv \, =$, then the space $I_i \times I_j$ is divided by the line $\sum_{i=1}^{n} a_i s_i(t) + \sum_{i=1}^{n} b_i s_i(t^\ast) = 0$ into two subsets $S^+_{i, j}$ and $S^-_{i, j}$.
%\end{enumerate}

Imagine we are in the situation from Theorem \ref{lemma}, but working with a noisy signal. We cannot be sure about the border of the set $S^+_{i, j}$ due to the error $\epsilon$. The condition $\sum_{i=1}^{n} a_i s_i(t) + \sum_{i=1}^{n} b_i s_i(t^\ast) = b$ defining points on the border of $S^+_{i, j}$ could be easily broken by changing the values of the signal $s$ by arbitrary small values from $(0, \epsilon)$. For this reason, we will ignore these border points and count them as they were in $S^+_{i, j}$ or in $S^-_{i, j}$ depending on what would be computationally easier. As a consequence, in the first case in Theorem \ref{lemma}, $S^+_{i, j} = \emptyset$, and in the second case, it does not matter if $\sim{}\equiv{}<$ or $\sim{}\equiv{}\leq$.

As a result of this simplification, the set of allowed operators in atomic predicates is restricted to $\lbrace <, > \rbrace$. The remaining three operators lost their sense. With signals burden with some error $\epsilon > 0$ the satisfaction of atomic predicates can be meaningfully determined only for time instants inside the satisfaction set. 
%(Figure \ref{obr_signal}). 
Hence the satisfaction of a formula using the operator $=$, e.g., $x = b$ or $x = y + 2$, cannot be determined. Every reference to a precise value has to be replaced by a sufficiently large interval, e.g., formula $x = b$ can be replaced by $x \geq (b - \delta) \wedge x \leq (b + \delta)$ for some $\delta > \epsilon / 2$.

Based on these assumptions, monitoring can be solved approximately. For every pair of intervals $I_i, I_j \in\intervals$, the satisfaction set on this area can be described by a single convex polygon $S^+_{i,j}$. For the whole satisfaction set $S_{\mu,s}$ we get $S_{\mu,s} = \bigcup_{i, j \in\{1,...,m\}} S^+_{i,j}$. The problem of construction of satisfaction sets for composite formulae from Theorem \ref{monitoring_STL*} is reduced to operations with sets of convex polytopes in plane for which efficient algorithms exist~\cite{GeometricAlgorithms}. 
 
\begin{algorithm}[Approximative monitoring algorithm for piecewise linear signals]
\label{algorithm1}

\end{algorithm}
\smallskip
\textbf{Input:} A piecewise linear signal $s$ and an STL* formula $\varphi$.\\
\textbf{Output:} Answer to the question $s \models \varphi$.\\

Algorithm inductively constructs the satisfaction set $S_{\varphi}$:\\

All the computations are performed on parts of the signal $s$ of sufficient length. This can be computed according to \eqref{sufficientSignalLength}. If the signal does not have the sufficient length, the answer returned by the algorithm might be wrong.

\begin{enumerate}

	\item $S_{\mu} = \bigcup_{i, j \in\{1,...,m\}} S^+_{i,j}$. The computation of the satisfaction sets $S^+_{i,j}$ for individual intervals is performed according to Theorem \ref{lemma}, but without determining the satisfaction of the borders (as was justified in this section).

	\item $S_{\neg \varphi} = (\intervals \times \intervals) \setminus S_{\varphi}$. The result can be computed as a simple Boolean operation on polygons. It is necessary to ensure that the resulting set consists only of convex polygons, which could be achieved for example by triangulation~\cite{GeometricAlgorithms}. 
	
	\item $S_{\varphi_1 \vee \varphi_2} = S_{\varphi_1} \cup S_{\varphi_2}$. Again a Boolean polygonal operation. 

	\item $S_{\ast \varphi} = \lbrace (t, t^\ast) \in \mathbb{R}_0^+ \times \mathbb{R}_0^+ \mid (t, t) \in S_{\varphi}, t^\ast \in \mathbb{R}_0^+ \rbrace$. The satisfaction set for the freeze operation can be computed by: (1) finding an intersection between the line $t^* = t$ and the satisfaction set $S_{\varphi}$ (black line segments in Figure~\ref{figFreezeSS-a}), (2) substituting values in the second component by the whole axis $t^\ast$, i.e., making a projection to the first component $t$ and then a Cartesian product with the axis $t^\ast$ (Figure \ref{figFreezeSS-b}).

\begin{figure}[h]
   \centering  
	\subfloat[][]{\includegraphics[scale=0.4]{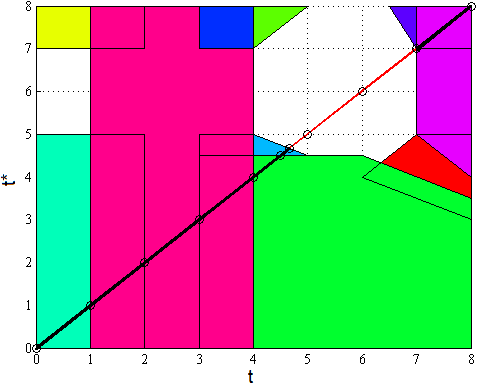}
			\label{figFreezeSS-a}}
    \subfloat[][]{\includegraphics[scale=0.4]{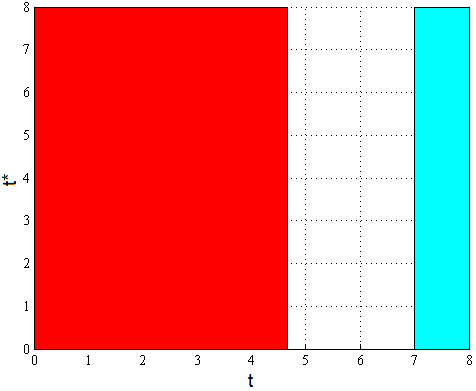}
			\label{figFreezeSS-b}}
    	
      	\caption{Example of computing satisfaction set for the formula $\ast \varphi$. a) The satisfaction set represented as a set of (overlapping) convex polygons (different colors are used only for depicting the fact that the set consists of convex polygons) and the line $t = t^\ast$. The regions of intersection of the line and the satisfaction set are depicted in black color. b) Resulting set $S_{\ast \varphi}$ obtained by projecting the intervals of intersection to the axis $t$ and making a Cartesian product with the axis $t^\ast$.}
	\label{figFreezeSS}    
    
\end{figure}

	\item \label{step} $S_{\varphi_1 \until{a}{b} \varphi_2} = \lbrace (t, t^\ast) \in S_{\varphi_1} \mid \exists t' \in [t + a, t + b]: (t', t^\ast) \in S_{\varphi_2} \wedge \forall t'' \in [t, t'] : (t'', t^\ast) \in S_{\varphi_1}\rbrace$. To be more illustrative, we explain this part of the algorithm from the geometrical point of view, identifying values $t$ with the horizontal axis and values $t^\ast$ with the vertical axis.

\begin{figure}[h]
   \centering  
	\subfloat[][$S_{\varphi_1}$]{\includegraphics[scale=0.4]{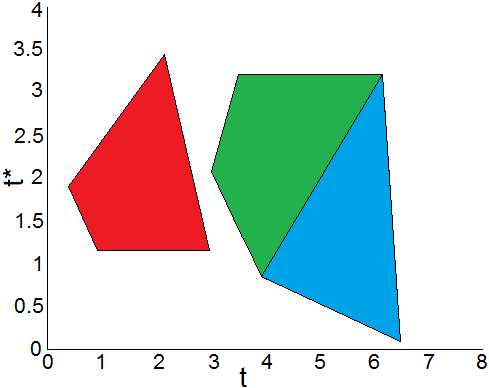}
			\label{figAlgorithm1-a}}
    \subfloat[][$S_{\varphi_2}$]{\includegraphics[scale=0.4]{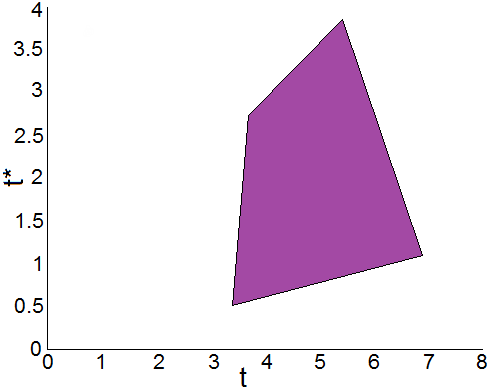}
			\label{figAlgorithm1-b}}
    \subfloat[][$S_{\varphi_1} \cap S_{\varphi_2}$]{\includegraphics[scale=0.4]{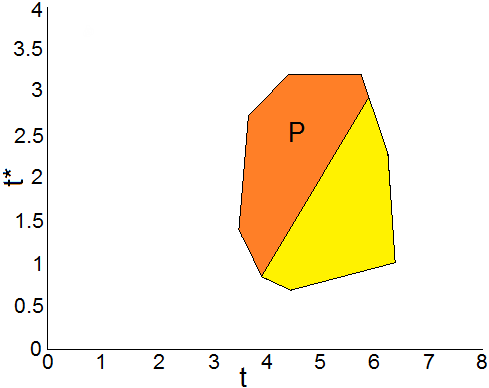}
			\label{figAlgorithm1-c}}
      	\caption{First step of computation of $S_{\varphi_1 \untilLTL_{[a, b]} \varphi_2}$ is to find the intersection $S_{\varphi_1} \cap S_{\varphi_2}$.}
	\label{figAlgorithm1}    
    
\end{figure}	

\begin{figure}[h]
   \centering  
	\subfloat[][$P \cup S_{\varphi_1}$]{\includegraphics[scale=0.4]{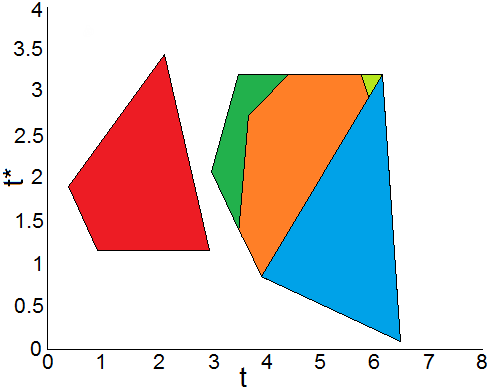}
			\label{figAlgorithm2-a}}
    \subfloat[][ordered vertices]{\includegraphics[scale=0.4]{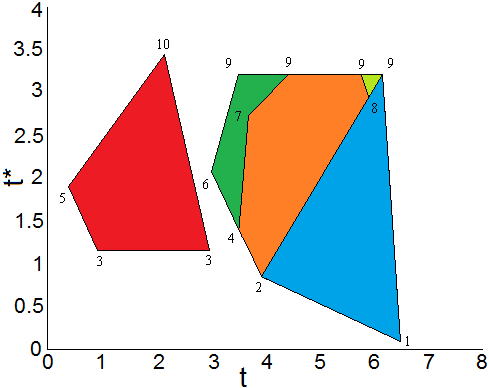}
			\label{figAlgorithm2-b}}
    \subfloat[][rectangular polygons (stripes) \\$R_i, i = 1, \ldots, n - 1$]{\includegraphics[scale=0.4]{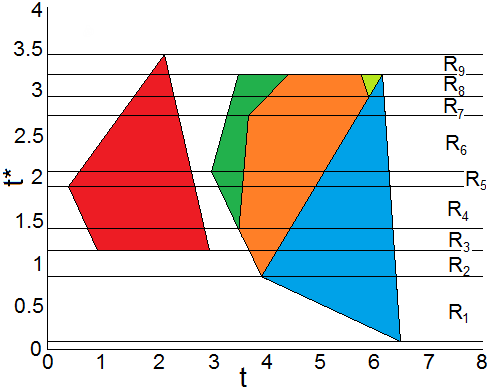}
			\label{figAlgorithm2-c}}
      	\caption{Next, the space is divided into horizontal stripes. The task is solved in each stripe separately.}
	\label{figAlgorithm2}    
    
\end{figure}
	
	For each convex polygon $P \in S_{\varphi_1} \cap S_{\varphi_2}$ (Figure~\ref{figAlgorithm1-c}) the following procedure is performed:
		 \begin{enumerate}
			\item Vertices of polygons $\lbrace P \rbrace \cup S_{\varphi_1}$ are increasingly ordered by the second component $t^\ast$ (duplicates can be removed). Denote the ordered set $V \df (V_1, V_2, \ldots V_n)$ (Figures~\ref{figAlgorithm2-a},~\ref{figAlgorithm2-b});
			\item For $i = 1, \ldots, n-1$ every neighbouring pair of vertices $V_i = (t_{i}, t^\ast_{i}), V_{i+1} = (t_{i+1}, t^\ast_{i+1}) \in V$ specifies a rectangular polygon (stripe): $$R_{i} \df \lbrace (t, t^\ast) \in \mathbb{R}_0^+ \times [t_i^\ast, t_{i+1}^\ast] \rbrace$$ (Figure~\ref{figAlgorithm2-c}). 
			$R_{i}$ in a form of line segment is considered empty because we do not care about the border points.
			%(We could use $>$ instead of $\geq$ because we do not care about border points. For the same reason, $R_{i}$ in a form of line segment is considered empty.) 
					
			\item The task is solved in every stripe $R_{i}$ separately. For every nonempty $R_{i}$ denote $P_{i} \df R_{i} \cap P$ and $S_{i} \df R_{i} \cap S_{\varphi_1}$. Due to the way of construction of $R_{i}$, all polygons in $S_{i}$ and the single polygon $P_{i}$ have all their vertices on the upper ($t^\ast = t^\ast_{i+1}$) or the lower ($t^\ast = t^\ast_{i}$) border of the stripe $R_i$ (Figure~\ref{figAlgorithm3-a}). In fact, these polygons can take only the shape of a triangle or a trapezoid (Figure \ref{figStripe}).\\
			
\begin{figure}[h]
   \centering  
	\subfloat[][the stripe $R_4$]{\includegraphics[scale=0.4]{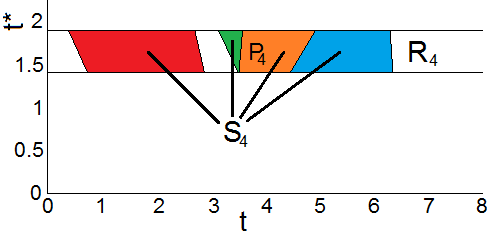}
			\label{figAlgorithm3-a}}
    \subfloat[][points on the left side of the line $\overrightarrow{ul}$]{\includegraphics[scale=0.4]{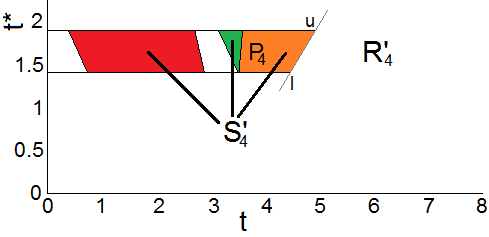}
			\label{figAlgorithm3-b}}
    \subfloat[][$A_4 = $ the rightmost polygon in $S'_{4}$]{\includegraphics[scale=0.4]{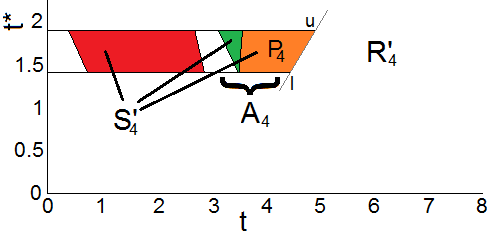}
			\label{figAlgorithm3-c}}
      	\caption{The potential polygons are identified. (The solution cannot be placed further in time than the property $\varphi_2$. Moreover, the property $\varphi_1$ must continuously hold at all time points until $\varphi_2$ is satisfied.)}
	\label{figAlgorithm3}    
    
\end{figure}				
\enlargethispage*{5mm}			
Now we get rid of the polygons lying on the right side of $P_{i}$ because their points cannot be in the solution (they represent the time past the event $\varphi_2$). To do so, we isolate the upper and lower rightmost vertices of the polygon $P_{i}$. Denote them as $u = (u_t, u_{t^\ast})$ and $l = (l_t, l_{t^\ast})$ (Figure~\ref{figAlgorithm3-b}). We get a new area:

$$R'_{i} \df \lbrace (t, t^\ast) \in R_{i} \mid (t, t^\ast) \textnormal{ \small lies on the left side of the line } \overrightarrow{ul} \rbrace.$$

 The solution can lie only in this area, so we can restrict $S_{i}$ to $S'_{i} \df R'_{i} \cap S_{i}$. 
			
\begin{figure}[h]
	\begin{center}
    	\includegraphics[scale=0.35]{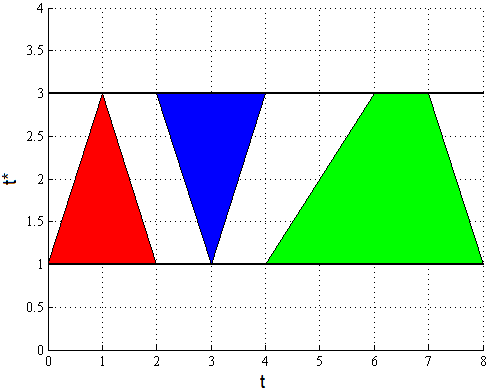}
      	\caption{An example of a stripe and all possible shapes of polygons inside a stripe.}
      	\label{figStripe}
    \end{center}
\end{figure}

			\item All the adjoining polygons (sharing an edge) in $S'_{i}$ are connected to form maximal seamless convex polygons (e.g., the green and orange polygons in Figure~\ref{figAlgorithm3-b}).  
			
			\item Let $A_{i}$ be the rightmost polygon in $S'_{i}$ (Figure~\ref{figAlgorithm3-c}) (Even after connecting some polygons in $S'_{i}$ their shape can still be only triangular or trapezoidal (Figure \ref{figStripe}) so there is only one rightmost polygon in $S'_{i}$). Then $A'_{i} = A_{i} \cap (P_{i} \ominus (b, a))$ is the solution for the stripe $R_{i}$ (Figure~\ref{figAlgorithm4}).\\
The operation $\ominus$ is defined as $A \ominus (a, b) \df \lbrace (x, y) \in \mathbb{R} \times \mathbb{R} \mid \exists c \in (a, b): (x + c, y) \in A \rbrace$ which is equal to $A \oplus (-a, -b)$, where $\oplus$ is the Minkowski sum.
			
			\item The final satisfaction set is given by $S_{\varphi_1 \until{a}{b} \varphi_2} = \bigcup_{i=1}^{n-1} A'_{i}$.   
			
\begin{figure}[h]
   \centering  
	\subfloat[][$P_4$ before application of the operator $\ominus$]{\includegraphics[scale=0.4]{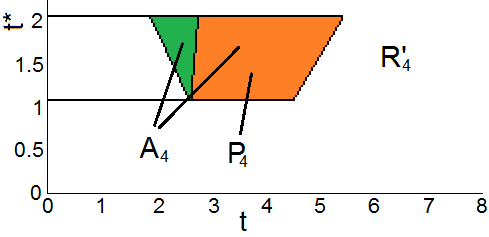}
			\label{figAlgorithm4-a}}
    \subfloat[][the shifted polygon $P_4 \ominus (2.5, 2.5)$]{\includegraphics[scale=0.4]{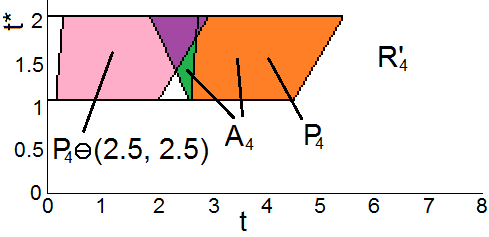}
			\label{figAlgorithm4-b}}
    \subfloat[][$A'_4 = A_4 \cap (P_4 \ominus (2.5, 2.5))$, the final solution in the stripe $R'_4$]{\includegraphics[scale=0.4]{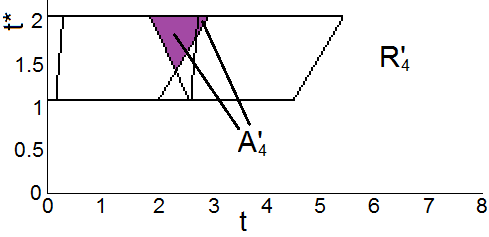}
			\label{figAlgorithm4-c}}
      	\caption{Last step is to consider the interval $[a, b]$ bounding the operator $\untilLTL$.}
	\label{figAlgorithm4}    
    
\end{figure}			
			
 \end{enumerate}

	\item The result $s \models \varphi$ which is equivalent to $(0, 0) \in S_\varphi$ is returned.

\end{enumerate}

\begin{remark}

The satisfaction set for temporal operator future $\future{a}{b} \varphi \equiv \mathsf{true} \until{a}{b} \varphi$ (and hence $\globally{a}{b} \varphi$) can be computed more easily than according to the step \ref{step}. It can be obtained directly as $S_{\future{a}{b} \varphi} = S_{\varphi} \ominus (b, a)$.

\end{remark}

%\end{algorithm}	

\subsection{Time and space complexity}
\enlargethispage{8mm}
Time and space complexity of all steps of the Algorithm \ref{algorithm1} is proportional to the number of polygons they are working with. Number of polygons generated in step 1 is at most $n^2$ for each atomic predicate, where $n$ is the number of intervals on which the signal is defined.

The number of polygons does not asymptotically change during the computation of operations described in other steps of the algorithm and all these functions do not work slower than $O(n^2)$ (see~\cite{GeometricAlgorithms} for more details). The output polygons also keep small number of vertices during the process.

The upper bound on the total computation time is therefore $O(kn^4)$, where $n$ is the number of intervals on which the signal is defined and $k$ is the size of the investigated formula.

\section{Evaluation and Case Study}
\label{sec:casestudy}

A prototype of the approximative monitoring Algorithm \ref{algorithm1} was implemented in Matlab (version R2011b). The Multi-Parametric Toolbox (MPT) package~\cite{mpt} (version 2.6.3) was used for the polygonal operations. We have not focused on efficiency of the implemented algorithm, our goal was to prove the concept of the presented algorithm. Detailed description of the implementation, results of the experiments and performance analysis can be found in~\cite{Dluhos}.

%In this section we demonstrate a typical use of STL* and the monitoring algorithm. The main contribution is the possibility to express nontrivial properties of continuous signals in combination with ability to automatically test them. Imagine we have a model of some system and we want to find the parameters of the model such that its behaviour meets desired requirements. One way would be to perform a qualitative analysis of the underlying ODEs. But it is not always feasible, especially for more complex systems. Also, the investigated model does not have to be in the form of differential equations. We might want to study parameters of systems defined in other formalisms or even existing only in the real world.

%In these cases we have to relay on manual verification of satisfaction of desired properties. However, with the tool for algorithmic monitoring we can run a lot of experiments with different parameters and automatically verify which combination satisfies our needs. Let us demonstrate this on a case study.

\begin{figure}[h]
	\begin{center}
    	\includegraphics[scale=0.30]{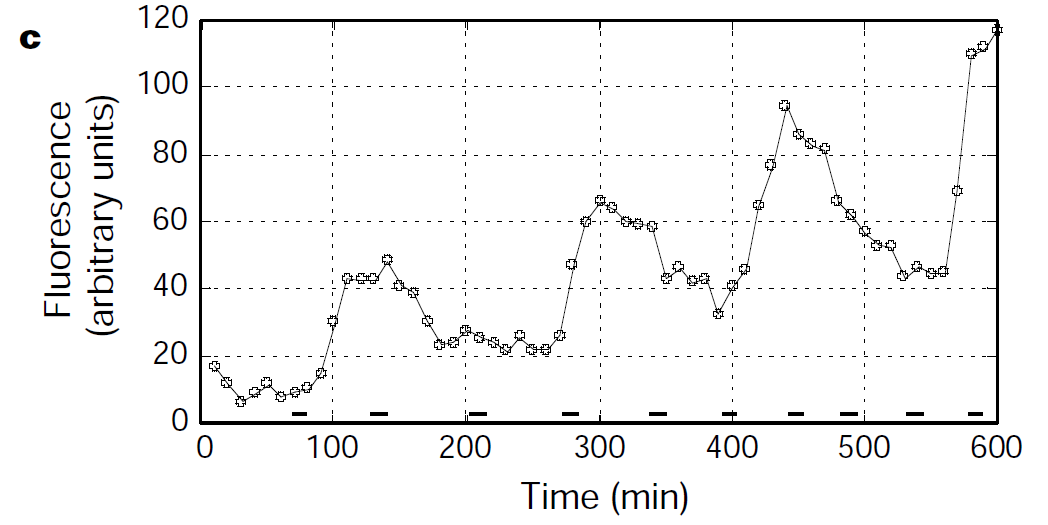}
      		\caption{The fluorescent intensity of a colony of \textit{Escherichia coli} containing the repressilator. The intensity is increasing overall due to the increasing number of individuals in the colony. The period of oscillations is lower than the time between cell division, but the whole repressilator is transmitted from generation to generation. Figure was taken from~\cite{Elowitz_Repressilator}.}
      	\label{figRepress2}
    \end{center}
\end{figure}

We have studied a biological system of three transcriptional repressors which was designed and built into bacteria \textit{Escherichia coli}~\cite{Elowitz_Repressilator}. Concentrations of involved proteins are periodically oscillating which causes periodic production of a green fluorescent protein. Intensity of fluorescence of this protein, which can be measured, gives evidence of the ongoing activity of the network (Figure \ref{figRepress2}).

%    We have chosen this system because it shows nontrivial oscillatory behaviour, it is simple enough to perform the computations with the prototype algorithm and it is a well-studied system. The goal of this case study is primarily to demonstrate the ways of usage of STL*.

\enlargethispage{8mm}

The network was modelled as a system of six differential equations: 
\begin{equation}
\small
	\label{eqRepress}
	\begin{gathered}
		\begin{array}{lclrcr}
			\dot{m_1} & = & -m_1 + \frac{\alpha}{1 + p^n_3} + \alpha_0; & \dot{p_1} & = & -\beta(p_1 - m_1);\\
			\dot{m_2} & = & -m_2 + \frac{\alpha}{1 + p^n_1} + \alpha_0; & \dot{p_2} & = & -\beta(p_2 - m_2);\\
			\dot{m_3} & = & -m_3 + \frac{\alpha}{1 + p^n_2} + \alpha_0; & \dot{p_3} & = & -\beta(p_3 - m_3).\\
		\end{array}
	\end{gathered}
\end{equation}\\

\begin{figure}[h]
	\begin{center}
    	\includegraphics[scale=0.45]{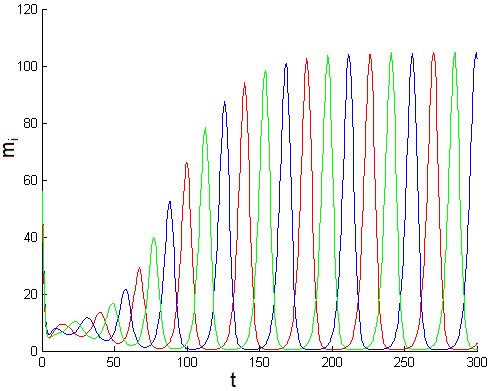}
      		\caption{A typical signal produced by the system \eqref{eqRepress}. Only variables $m_1, m_2$ and $m_3$ are depicted.}
      	\label{FigRepressProgress}
    \end{center}
\end{figure}

Where $\dot{m_i} \df \dfrac{dm_i}{dt}, \dot{p_i} \df \dfrac{dp_i}{dt}; m_1, m_2$ and $m_3$ correspond to activity of genes $lacl, tetR$ and $cl$; $p_1, p_2$ and $p_3$ are concentrations of corresponding proteins and $\alpha, \beta, \alpha_0, n$ are constants (a typical behaviour of the system is depicted in Figure~\ref{FigRepressProgress}).

The analysis of the system \eqref{eqRepress} and the estimation of parameters $\alpha, \beta, \alpha_0, n$ producing the oscillatory behaviour was originally done by means of manual qualitative analysis of ODEs~\cite{Elowitz_Repressilator}. However, we can perform the analysis automatically using the monitoring of STL* formulae. We can specify the desired oscillatory property by the formula:
\begin{equation}
	\label{formulaRep1}
	\varphi \df \globally{10}{190} \future{0}{50} \ax{(\future{1}{50} m_i^\ast < m_i) \wedge \future{1}{50} m_i^\ast > m_i)}
\end{equation}
and test the satisfaction for different values of parameters on runs of the length at least 300 minutes. The expected period has to be lower than 50 minutes. We start the testing 10 minutes later after the beginning to avoid the initial swing and end the testing 50 minutes before the end of the measurement because the signal might be cut in the middle of a period.

\begin{figure}[ht]
	\centering
		\subfloat[][$\alpha = 400, \beta = 0.2, \alpha_0 = 0.2, n = 2$]{\includegraphics[scale=0.45]{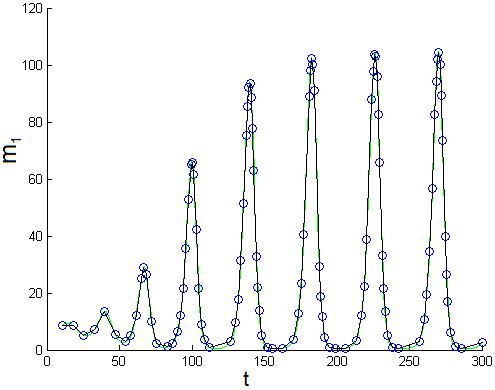}
			\label{figRepress3-a}}
		\subfloat[][$\alpha = 400, \beta = 0.2, \alpha_0 = 2, n = 2$]{\includegraphics[scale=0.45]{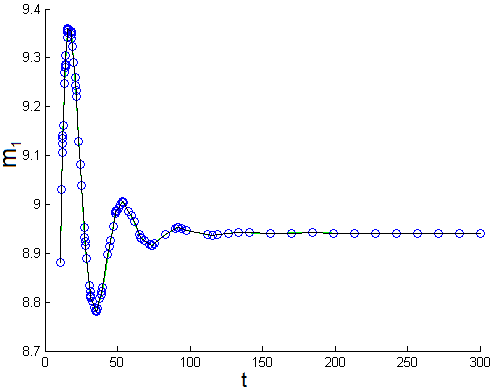}
			\label{figRepress3-b}}
	\caption{Sample runs of the system \eqref{eqRepress} for different sets of parameters. The initial values were $m_1 = 0.1, m_2 = 0.3, m_3 = 0.2, p_1 = 0.2, p_2 = 0.1, p_3 = 0.3$. Only the values of $m_1$ are depicted.}
	\label{figRepress3}
\end{figure}

Formula \eqref{formulaRep1} is satisfied by both runs in Figure \ref{figRepress3}. To avoid the case of damped oscillations, we can add the formula:
\begin{equation}
	\label{formulaRep2}
	\psi \df \globally{10}{200} \ax{(\future{1}{50} m_i^\ast \leq m_i)}.
\end{equation}

It ensures that the values reached in each period are not decreasing. By connecting formulae \eqref{formulaRep1} and \eqref{formulaRep2} we get the formula $\varphi \wedge \psi$. It does express the desired oscillatory behaviour in Figure \ref{figRepress3-a} and it is not satisfied by signals of the type depicted in Figure \ref{figRepress3-b}.   
  
\begin{figure}[h!t]
	\begin{center}
    	\includegraphics[scale=0.45]{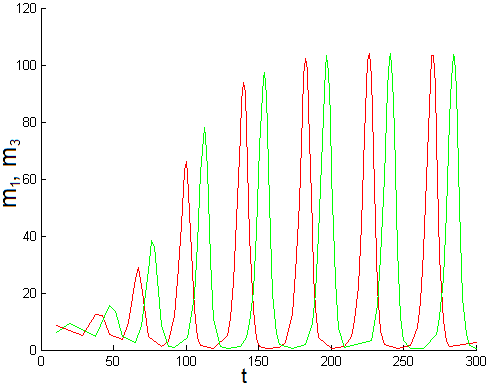}
      		\caption{Sample run of the system \eqref{eqRepress} satisfying the formula \eqref{formulaRep3}. The parameters were $\alpha = 400, \beta = 0.2, \alpha_0 = 0.2, n = 2$ and the initial values $m_1 = 0.1, m_2 = 0.3, m_3 = 0.2, p_1 = 0.2, p_2 = 0.1, p_3 = 0.3$. Depicted variables are $m_1$ -- red and $m_3$ -- green.}
      	\label{figRepress4}
    \end{center}
\end{figure}
  
Another property of the repressilator can be expressed by the formula:
\begin{equation}
	\label{formulaRep3}
	\globally{0}{270} \ax{(\future{0}{30} (m_1^\ast + 1 > m_3 \wedge m_1^\ast - 1 < m_3)},
\end{equation}\\
which means that \textit{"the values of $m_1$ precede the values of $m_3$ in such sense that for every value of $m_1$, $m_3$ reaches similar value in short time after $m_1$ did"} (Figure \ref{figRepress4}).

Full estimation of parameters could not be performed due to high time demands of the implemented monitoring algorithm~\cite{Dluhos}. A single run of the monitoring algorithm for the formulae \eqref{formulaRep1}, \eqref{formulaRep2} or \eqref{formulaRep3} over a signal sampled by 80 points took several hours on a regular PC. Reduction of the number of points describing the signal would lead to excessive loss of information. We have found out that the bottleneck in efficiency of the prototype implementation lies in polyhedral operations performed by MPT. Since we work in 2D and need only a small subset of operations, we believe that our algorithm can be significantly accelerated when an optimal implementation is employed. Here we focused on demonstrating the applicability while leaving efficiency for future work.

Different task would be to ensure that the behaviour of concentrations of the fluorescent protein in the bacteria (Figure \ref{figRepress2}) is in correspondence with the model. While it could be done only manually in~\cite{Elowitz_Repressilator}, if we had the data, we would be able to test the properties of the measurements identically as the properties of the signals produced by ODEs.

\enlargethispage*{15mm}

\vspace{-1mm}
\section{Conclusion}
\vspace{-1mm}

We have proposed an extended Signal Temporal Logic STL$^*$ motivated by the need to express properties of biological dynamical systems in a detail sufficient to distinguish different shapes of oscillation. We have provided a monitoring algorithm that approximately computes the truth value of an STL$^*$ formula for a given continuous (piece-wise linear) signal. The method has been prototyped in Matlab and the results achieved on a case study of oscillatory behaviour of the repressilator showed that the method satisfactorily works for signals generated by numerical simulation. However, the employed library MPT for polyhedral operations appears to be not efficient enough to satisfy the needs of practical usage. 

For future work on the practical side, we plan to implement more efficient algorithms for the specific polyhedral operations we use in the monitoring procedure. %Since we primarily need to efficiently compute 1D and 2D operations 
Another straightforward direction of future development is lifting of the robustness measure~\cite{Donze_Robustness} to the extended logic. 

A very inspiring work is~\cite{ATVA} where Time-Frequency Logic (TFL) is defined by shifting STL semantics to frequency domain. TFL provides another way to express permanent oscillations. However, non-pure oscillatory behaviour such as damped oscillations require specific elaboration in the time-domain. Joining TFL semantics with the STL$^*$ concept of real-value freezing can be an interesting step further.

\bibliographystyle{eptcs} % or whatever you prefer
\bibliography{hsb2012}

%\newpage

% \begin{appendix}
% \input{6_appendix_compmod.tex}
% \end{appendix}

\end{document}